# Using Geoelectrons to Search for Velocity-Dependent Spin-Spin Interactions

L.R. Hunter and D. Ang

*Physics Department, Amherst College, Amherst, MA 01002*

We use the recently developed model of the electron spins within the Earth to investigate all of the six possible long-range velocity-dependent spin-spin interactions associated with the exchange of an intermediate vector boson. Several laboratory experiments have established upper limits on the energy associated with various fermion-spin orientations relative to the Earth. We combine the results from three of these experiments with the geoelectron-spin model to obtain bounds on the velocity-dependent interactions that couple electron spin to the spins of electrons, neutrons and protons. Five of the six possible potentials investigated were previously unbounded. The bound achieved on $V_8$ is about 30 orders of magnitude more restrictive in the long-range limit than the only previously established constraint.

Recently, there has been renewed interest in exploring possible anomalous spin-spin interactions mediated by new particles [1-7]. Observation of such an interaction would constitute the discovery of a new force in nature and suggest physics beyond the standard model of particle physics. Non electromagnetic spin-spin forces created through the exchange of a scalar boson (like the axion) were first suggested by Moody and Wilcek [8]. Dobrescu and Mocioiu enumerated nine possible spin-spin interactions associated with the exchange of a vector boson (like the *z'*) that are compatible with rotational invariance [9]. Stringent limits have now been placed on the three velocity-independent interactions both at long range [1, 4, 5, 10 ] and at atomic scales [2, 3, 7]. The remaining six interactions (numbered as in Ref. [9]) depend not only on the spins ($\hat{\sigma}$) and relative positions (**r**) of the two fermions, but also on their relative velocity (**v**).

$$V_{6,7} = -\frac{\hbar}{8\pi c^2}\left(\frac{g_V^1 g_A^2}{2M_1} + \frac{g_A^1 g_V^2}{2M_2}\right)\left[(\hat{\sigma}_1 \cdot \mathbf{v})(\hat{\sigma}_2 \cdot \hat{\mathbf{r}}) \pm (\hat{\sigma}_1 \cdot \hat{\mathbf{r}})(\hat{\sigma}_2 \cdot \mathbf{v})\right]\left(1+\frac{r}{\lambda}\right)\frac{e^{-r/\lambda}}{r^2} \quad (1)$$

$$V_8 = \frac{g_A^1 g_A^2}{4\pi c^2}\left[(\hat{\sigma}_1 \cdot \mathbf{v})(\hat{\sigma}_2 \cdot \mathbf{v})\right]\frac{e^{-r/\lambda}}{r} \quad (2)$$

$$V_{14} = \frac{g_A^1 g_A^2}{4\pi c}\left[(\hat{\sigma}_1 \times \hat{\sigma}_2) \cdot \mathbf{v}\right]\frac{e^{-r/\lambda}}{r} \quad (3)$$

$$V_{15} = -\frac{g_V^1 g_V^2 \hbar^2}{8\pi c^3 M_1 M_2}\left[(\hat{\sigma}_1 \cdot (\mathbf{v} \times \hat{\mathbf{r}}))(\hat{\sigma}_2 \cdot \hat{\mathbf{r}}) + (\hat{\sigma}_1 \cdot \hat{\mathbf{r}})(\hat{\sigma}_2 \cdot (\mathbf{v} \times \hat{\mathbf{r}}))\right]\left(3 + \frac{3r}{\lambda} + \frac{r^2}{\lambda^2}\right)\frac{e^{-r/\lambda}}{r^3} \quad (4)$$

$$V_{16} = -\frac{\hbar}{8\pi c^2}\left(\frac{g_V^1 g_A^2}{2M_1} + \frac{g_A^1 g_V^2}{2M_2}\right)\left[(\hat{\sigma}_1 \cdot (\mathbf{v} \times \hat{\mathbf{r}}))(\hat{\sigma}_2 \cdot \mathbf{v}) + (\hat{\sigma}_1 \cdot \mathbf{v})(\hat{\sigma}_2 \cdot (\mathbf{v} \times \hat{\mathbf{r}}))\right]\left(1+\frac{r}{\lambda}\right)\frac{e^{-r/\lambda}}{r^2} \quad (5)$$

In the above equations $g$ denotes the vector (V) or axial (A) coupling constants of fermions 1 or 2 with mass $M$. The interaction range of the force is denoted by $\lambda = \hbar/m_{z'}c$ where $m_{z'}$ is the mass of the intermediate vector boson, $\hbar$ is Planck's constant ($h$) divided by $2\pi$ and $c$ is the speed of light. We note that the potentials $V_6$ and $V_7$ violate time-reversal symmetry (T) while $V_{16}$ violates parity (P). The potentials $V_{14}$ and $V_{15}$ violate both T and P. Only the potential $V_8$ is both T and P conserving.

Recent experiments specifically designed to search for long-range spin-spin interactions employ spin-polarized laboratory sources and look for interactions between these sources and the spins contained in either a nuclear magnetometer [4, 5] or an electron-spin polarized torsion pendulum [6, 10]. These experiments are unable to constrain the velocity-dependent interactions because there is on average no relative velocity between the source and detection spins. Of the six possible velocity-dependent potentials, only $V_8$ for interactions between protons and neutrons (p-n) has been previously constrained. This was accomplished by comparing the theoretical and experimental spin-exchange interaction between Na and He atoms [3]. In that analysis, the non-zero value of the average relative velocity squared of the colliding atoms was used to yield a well defined constraint.

A new approach to the study of anomalous long-range spin-spin interactions was recently reported [1]. In the presence of the geomagnetic field, some of the electrons within paramagnetic minerals within the Earth acquire a small spin polarization. The magnitude and direction of the induced geoelectron spins were calculated using the known strength of the geomagnetic field [11] and a model of the Earth's composition and temperature. These spin-polarized geoelectrons can interact (via the proposed anomalous spin-spin potentials) with the electrons or nucleons contained in spin-sensitive detectors. Such an interaction can induce an energy shift in the detection spins that depends on their orientation with respect to the Earth. The reversibility of these potentials with the reversal of the detector spin-orientation provides a critical experimental signature. Three experiments have established bounds on various orientation-dependent energy shifts ($\beta$). The bounds on the electron ($e$) energy when its spin is oriented North ($N$) and East ($E$) are derived from the Seattle (47.658° N, 122.3° W) spin-polarized torsion-pendulum experiment: $\hat{\beta}_N^e < 5.9 \times 10^{-21} eV$ and $\beta_E^e < 8 \times 10^{-22} eV$ [12]. The bounds on the neutron ($n$) and proton ($p$) orientation-dependent energy obtained from the Amherst (42.37° N, 72.53° W) experiments are $\hat{\beta}_N^n < 4.2 \cdot 10^{-21} eV$ and $\hat{\beta}_N^p < 4.3 \cdot 10^{-20} eV$ [13] and $\beta_E^n < 2.9 \times 10^{-21} eV$ and $\beta_E^p < 3 \times 10^{-20} eV$ [1]. Another Seattle experiment, initially intended to measure possible couplings between nuclear spin and gravity, yields a bound for proton and neutron spin orientations along the Earth's spin axis ($z$): $\hat{\beta}_z^n < 1.2 \times 10^{-20} eV$ and $\hat{\beta}_z^p < 1.8 \times 10^{-20} eV$ [14]. Here, we have followed the convention of using a hat over beta to indicate that a correction has been applied to correct for the Earth's gyroscopic frequency. All bounds quoted are two standard deviations (i.e. 95% confidence level).

In [1] bounds from these three experiments were used to extract constraints on velocity-independent spin-spin interactions. Here we suggest that these geoelectrons can also serve as a source for the investigation of velocity-dependent spin interactions. Indeed, the Earth's rotation ($\mathbf{\Omega} = 2\pi/(1\ sidereal\ day)\hat{\mathbf{z}}$) creates substantial relative velocities between the source geoelectrons and laboratory spins and it only remains to sum the velocity-dependent potential contributions over all of the Earth's spins.

In our calculation we use the geoelectron spin densities derived in [1]. These densities are derived assuming that the Earth's paramagnetism comes predominately from the unpaired *d*-shell electrons in the iron ions contained in the minerals of the Earth's mantle and crust. In the model, the magnetization of the Earth's core is assumed to be negligible as is predicted by density functional theory (DFT) calculations [15-18].

The velocities relevant to the calculation are illustrated in Figure 1. The integration required to sum over each of the various potentials (equations 1-5) is outlined in Fig. 2. The procedure is similar to that followed in [1] with the addition of the appropriate relative velocity terms for the geoelectrons. The integration is carried out in geocentric coordinates using Mathematica.

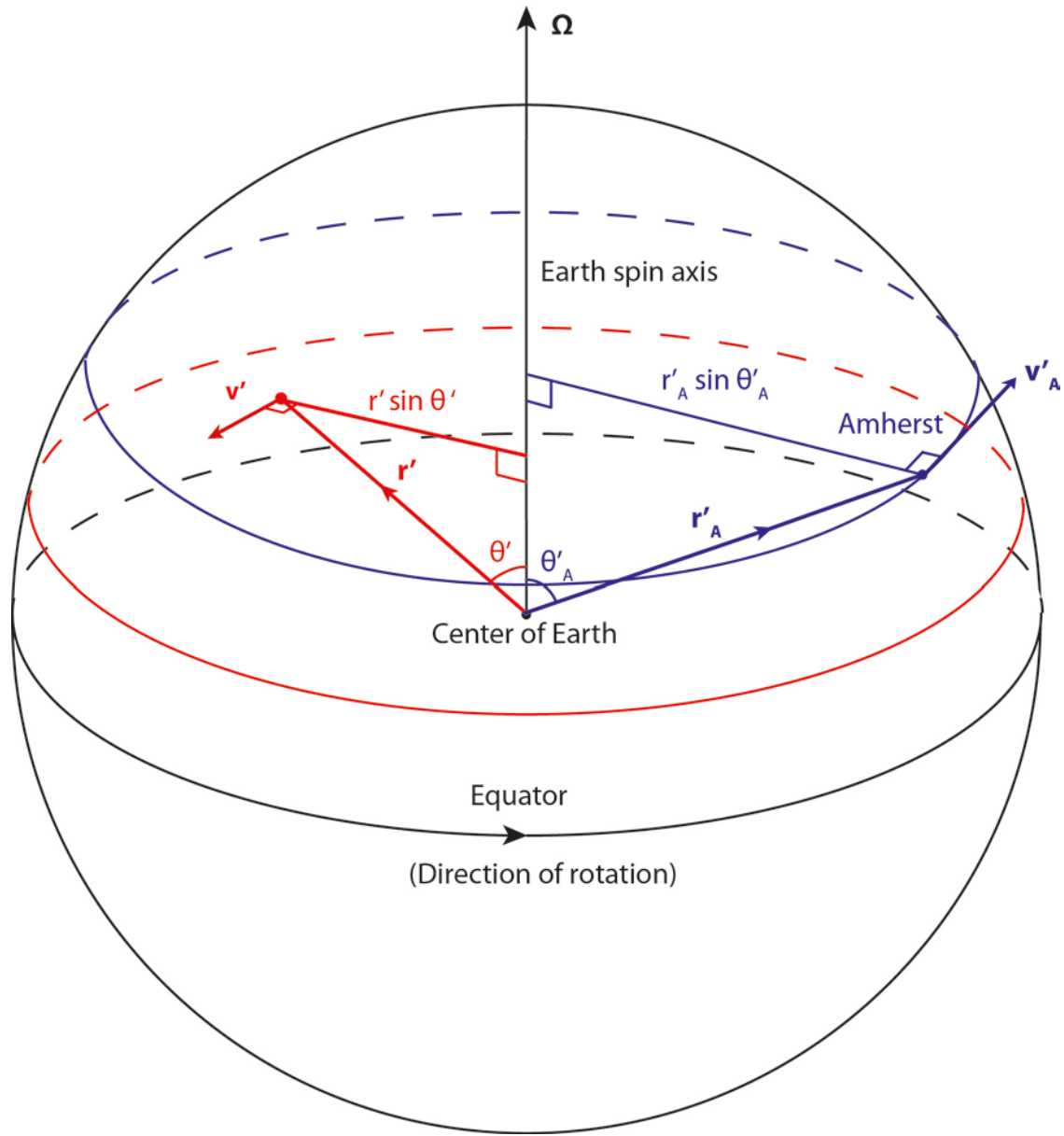

FIG. 1. The determination of the relative velocity between the spins. In order to evaluate the velocity dependent potentials (given by equations 1-5.) it is necessary to calculate $\mathbf{v} = \mathbf{v}'_A - \mathbf{v}'$, where $\mathbf{v}' = \mathbf{\Omega} \times \mathbf{r}'$ is the velocity of a geoelectron at position $\mathbf{r}'$ and $\mathbf{v}'_A = \mathbf{\Omega} \times \mathbf{r}'_A$ is the velocity of an electron in the detection apparatus in Amherst (or Seattle) at position $\mathbf{r}'_A$.

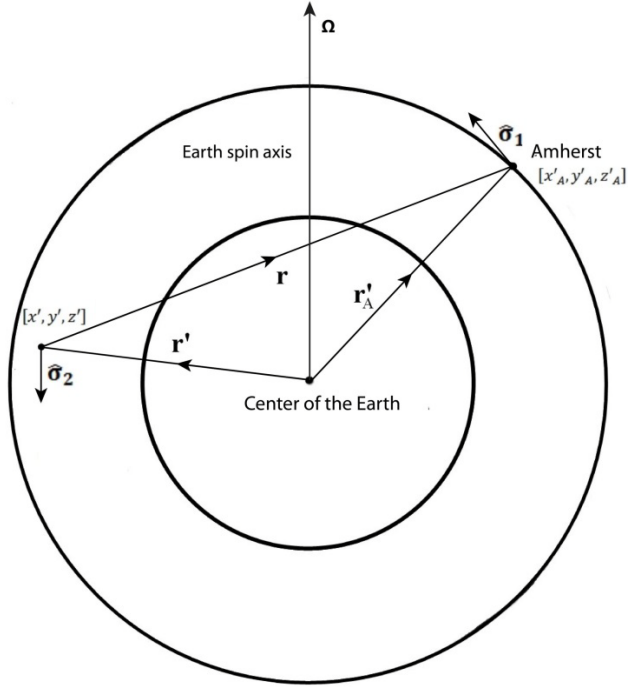

$\hat{\sigma}_1$ = spin sensitive direction of apparatus at Amherst

$\hat{\sigma}_2 = -\hat{\mathbf{B}}$

$\mathbf{r}'_A$ = Radial position vector of Amherst = $[x'_A, y'_A, z'_A]$

$\mathbf{r}'$ = Radial position vector of electron = $[x', y', z']$

$x' = r' \sin\theta' \sin\varphi'$

$y' = r' \sin\theta' \cos\varphi'$

$z' = r' \cos\theta'$

$\mathbf{r}(r', \theta', \varphi') = \mathbf{r}'_A - \mathbf{r}' = [(x'_A - x'), (y'_A - y'), (z'_A - z')]$

$\Omega = \dfrac{2\pi}{1 \text{ sidereal day}} \hat{\mathbf{z}}$

$\mathbf{v}'_A = \Omega \times \mathbf{r}'_A$

$\mathbf{v}' = \Omega \times \mathbf{r}'$

$\mathbf{v}(r', \theta', \varphi') = \mathbf{v}'_A - \mathbf{v}'$

$$V_{total} = \int_0^{2\pi} \int_0^{\pi} \int_{R_{CM}}^{R_S} r'^2 \sin\theta' * \rho(r') * \dfrac{2\mu_b B}{k_b T(r')} * V(\mathbf{r}(r', \theta', \varphi'), \mathbf{v}(r', \theta', \varphi'))\, dr'\, d\theta'\, d\varphi'$$

FIG. 2. - Details of the integration of the potential over the Earth volume. The integration is over all of the volume from the core-mantle boundary ($R_{CM}$) to the surface ($R_S$). The potential (given by one of Eqs. 1-5) is evaluated at $\mathbf{r} = \mathbf{r}'_A - \mathbf{r}'$ where the geoelectron location is described by the vector $\mathbf{r}'$ and the location of Amherst (or Seattle) is designated by the vector $\mathbf{r}'_A$. The geoelectron spin direction $\hat{\sigma}_2$ is assumed to be antiparallel to the Earth magnetic field $\mathbf{B}$ at $\mathbf{r}'$. The relative velocity of Amherst (or Seattle) to the geoelectron is given by $\mathbf{v} = \mathbf{v}'_A - \mathbf{v}'$ where $\mathbf{v}'_A = \Omega \times \mathbf{r}'_A$ and $\mathbf{v}' = \Omega \times \mathbf{r}'$ where $\Omega$ is the angular rotation vector of the Earth. The unit vector $\hat{\sigma}_1$ is the spin-sensitive direction of the apparatus in Amherst (or Seattle). The unpaired electron density, $\rho(r')$, and the temperature profile, $T(r')$, are taken from [1]. To establish the bounds on the coupling coefficients we require $V_{total}$ to be less than the energy bound established on the spin coupling energy ($\beta$) in the spin-sensitive direction by the various experiments.

For the velocity-independent potentials, all the geocentric spherical shells contribute to the integration with the same sign (c.f. Fig. S3 in [1]). Hence, making conservative assumptions about the iron densities produced valid upper bounds on the dimensionless coupling parameters. The geocentric spherical shells contribute with a uniform sign to the potentials $V_7$ and $V_8$ and hence the upper bounds obtained from this model remain valid for these two potentials (Figs. 3b and 3c).

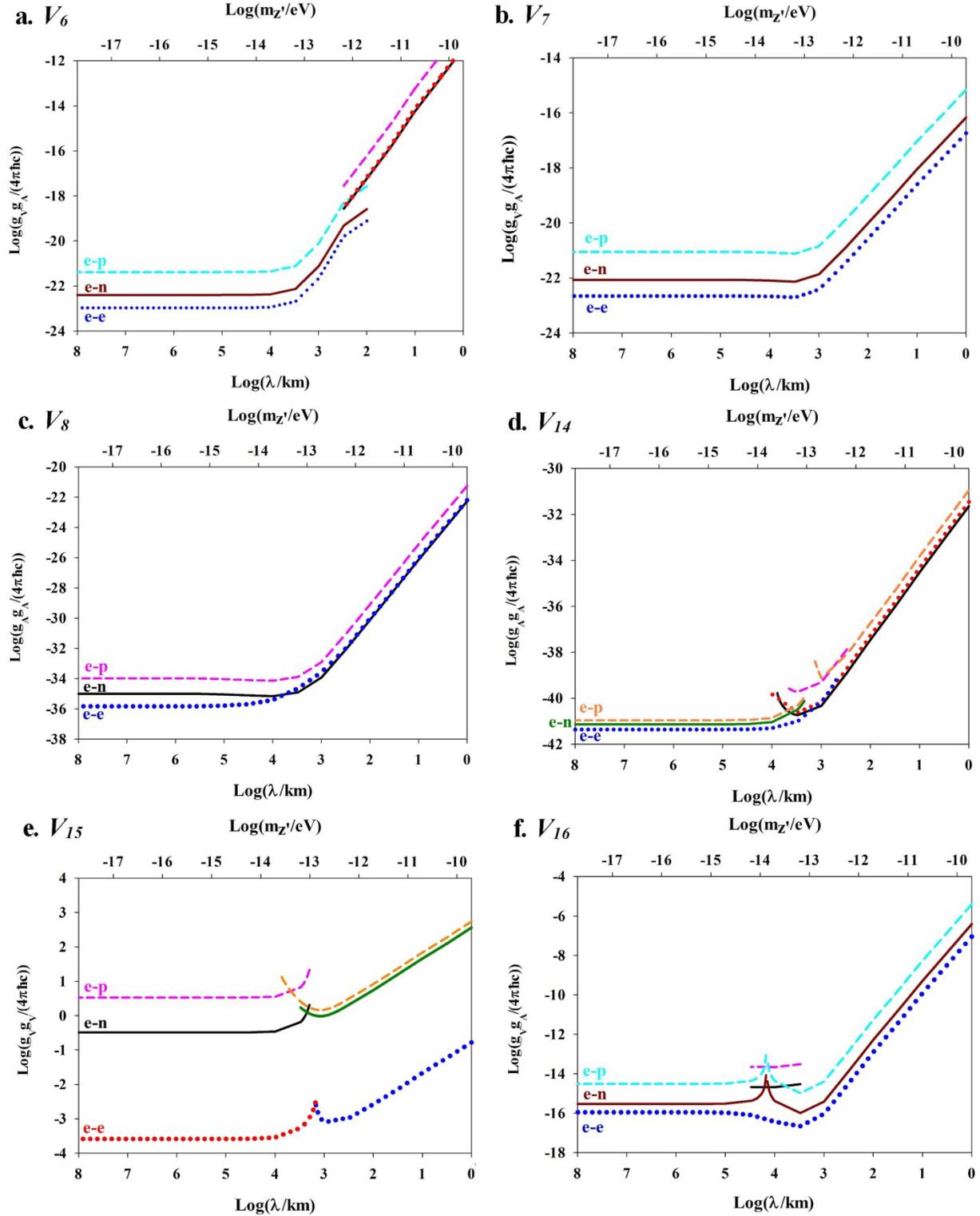

FIG. 3. Bounds on the long-range velocity dependent couplings. All regions above the lines are ruled out by experiment. The (e-e) bounds (dotted lines) are extracted from $\hat{\beta}_E^e$ (blue) and $\hat{\beta}_N^e$ (red). The (e-n) bounds (solid lines) are extracted from $\hat{\beta}_N^n$ (black), $\beta_E^n$ (brown), and $\beta_z^n$

(green). The (e-p) bounds (dashed lines) are derived from $\hat{\beta}_N^p$ (magenta), $\beta_E^p$ (cyan), and $\beta_z^p$ (orange). For vector-axial couplings (V-A) the first fermion from each labeled pair has the vector coupling while the second has the axial coupling. For (V-A) couplings the bounds for p-e and n-e couplings are $M_n/M_e \sim 1860$ times larger than the e-p and e-n bounds shown. (a) Vector-axial (V-A) couplings (Eq. 1 with the + sign) for $V_6$. (b) (V-A) couplings (Eq. 1 with the - sign) for $V_7$. (c) (A-A) couplings (Eq. 2) for $V_8$. (d) (A-A) couplings (Eq. 3) for $V_{14}$. (e) (V-V) couplings (Eq.4) for $V_{15}$. (f) (V-A) couplings (Eq. 5) for $V_{16}$.

For the potentials $V_6$, $V_{14}$, $V_{15}$, and $V_{16}$, different geocentric spherical shells can contribute to the interaction potential with different signs. This can result in substantial cancellation in the integration and for certain specific ranges the integral can completely cancel yielding no sensitivity whatsoever. For the ranges with high cancellation, modest changes in the electron density profile as a function of depth can significantly modify the resulting sensitivity. The model parameter that introduces the largest uncertainty in our result is the iron fraction contained in the lower mantle. While the average iron densities in the crust and upper mantle are reasonably well know, the densities in the lower mantle remain under debate [19]. In [1], a pyrolite model is assumed. This model posits that the iron fractions in the lower mantle are the same as those observed in the upper mantle. Alternative models suggest that the iron fractions in the lower mantle may be significantly larger. In order to take this model-dependent uncertainty into account, we have considered all configurations of the iron density between the pyrolite model and a rather extreme model where we allow the iron fraction to double throughout the lower mantle. To be conservative in our analysis, we report only the highest bound obtained from this entire array of plausible models. Because of the cancellation between the contributions of the lower and upper mantles, gaps appear in our bounds around some specific intermediate boson masses for some of the potentials. Fortunately, the same gaps do not appear for other orientations of the detector. As such, we have been able to "cap" these gaps with alternative bounds, albeit at a somewhat less stringent level.

The resulting bounds on the potential $V_6$ is shown in Fig. 3a. For this potential, orienting the spin detector East yields the most sensitive bounds for most ranges. However, the integration for this orientation exhibits a high degree of cancellation at short ranges (<300 km), rendering the bounds unreliable. Fortunately, when the detector is oriented North, no similar cancellation is observed in the integration. We therefore quote the less sensitive bounds from the North orientation at short range.

Each of the potentials $V_{14}$, $V_{15}$ and $V_{16}$ has a range where the integration exhibits a sign change for a particular orientation of the field. However, by combining the bounds from different experiments and different detector orientations we are able to place continuous bounds over the entire long-range region (Figs. 3d, 3e and 3f).

The bounds on the dimensionless couplings achieved here are the most restrictive on the axial-axial potentials that drop off as $1/r$, with the limits established on $V_{14}$ (which has one factor of $(v/c)$) significantly better than the limits on $V_8$ (which is proportional to $(v/c)^2$). At long range, the vector-axial potentials decay as $1/r^2$ and exhibit an intermediate sensitivity, with the bounds on the potentials that are linear in $v/c$ ($V_6$ and $V_7$) superior to the bounds on $V_{16}$, which is quadratic in $v/c$. The bounds on $V_{15}$, which drops off as $1/r^3$ at large $r$, are by far the least restrictive. The bounds for $V_{15}$ are also more susceptible to local inhomogeneities than the other potentials, and hence are somewhat less reliable, especially at short range.

We are unaware of any previously established bounds on five of the six potentials discussed here. For $V_8$, the long-range bounds reported here are about 30 orders of magnitude more restrictive than the previously reported (p-n) bound [3].

Recently, it has been suggested that longitudinal spin fluctuations may result in iron atoms within the core acquiring a local magnetic moment that could be as large as 1.3 Bohr magnetons [20, 21]. This proposal, which is in conflict with ab initio density functional calculations [15-18], is currently under debate. If this proposal is found to be correct our model will need to be extended into the core. The inclusion of such a large magnetization in the Earth's core would likely improve the bounds one could place on the very long-range (larger than 1,000 km) spin-spin interactions, albeit with much larger model-dependent uncertainties.

It is likely that the experimental bounds on the energies associated with the various fermion-spin directions ($\beta$) will improve significantly in the near future, further refining the limits that can be placed on spin-spin couplings using the spin-polarized geoelectron model. A new generation of the Amherst experiment hopes to achieve two orders of magnitude improvement [13]. A new $^{85}$Rb-$^{87}$Rb comagnetometer experiment, nearing completion at Cal State East Bay, will produce bounds on $\beta_z$ for the proton [22]. The Princeton $^3$He-K self-compensating magnetometer has demonstrated the highest intrinsic sensitivity and has recently been moved to the South pole [23]. With attention to calibration details, this experiment, which will be insensitive to the Earth's gyroscopic effects, should produce interesting values of $\beta$ for the neutron and proton.

We thank J.E. Gordon, S.K. Peck, J.-F. Lin, M. Romalis, J. Brown, M. Kozlov, W. Loinaz, K. Jagannathan, S. Maus, D. Hanneke and D. Alfe for useful conversations. This work has been supported by NSF grants PHY-0855465 and PHY-1205824. D. Ang would like to acknowledge support from the Schupf Scholarship program.

1   L. R. Hunter, J. E. Gordon, S. K. Peck, D. Ang, and J.-F. Lin, Science **339**, 5 (2013).
2   M. P. Ledbetter, M. V. Romalis, and D. F. J. Kimball, Physical Review Letters **110**, 040402 (2013).
3   D. F. Jackson Kimball, A. Boyd, and D. Budker, Physical Review A **82**, 062714 (2010).
4   G. Vasilakis, J. M. Brown, T. W. Kornack, and M. V. Romalis, Physical Review Letters **103**, 261801 (2009).
5   A. G. Glenday, C. E. Cramer, D. F. Phillips, and R. L. Walsworth, Physical Review Letters **101**, 261801 (2008).
6   C. E. Cramer, 2007.
7   S. G. Karshenboim, Physical Review A **83**, 062119 (2011).
8   J. E. Moody and F. Wilczek, Physical Review D **30**, 130 (1984).
9   B. A. Dobrescu and I. Mocioiu, Journal of High Energy Physics **11**, 005 (2006).
10  T. C. P. Chui and W.-T. Ni, Physical Review Letters **71**, 3247 (1993).
11  S. Maus, S. Macmillan, S. McLean, B. Hamilton, A. Thomson, M. Nair, and C. Rollins, edited by NESDIC/NGDC (NOAA, 2010).
12  B. R. Heckel, E. G. Adelberger, C. E. Cramer, T. S. Cook, S. Schlamminger, and U. Schmidt, Physical Review D **78**, 092006 (2008).
13  S. K. Peck, D. K. Kim, D. Stein, D. Orbaker, A. Foss, M. T. Hummon, and L. R. Hunter, Physical Review A **86**, 012109 (2012).
14  B. J. Venema, P. K. Majumder, S. K. Lamoreaux, B. R. Heckel, and E. N. Fortson, Physical Review Letters **68**, 135 (1992).
15  D. Alfè and M. J. Gillan, Physical Review B **58**, 8248 (1998).
16  P. Söderlind, J. A. Moriarty, and J. M. Wills, Physical Review B **53**, 14063 (1996).
17  L. Vocadlo, J. Brodholt, D. Alfe, G. D. Price, and M. J. Gillan, Geophys Res Lett **26**, 1231 (1999).
18  G. Steine-Neumann, R. E. Cohen, and L. Stixrude, Journal of Physics: Condensed Matter **16** (2004).
19  R. Jeanloz and E. Knittle, Phil. Trans. R. Soc. Lond. A **323**, 377 (1989).
20  L. V. Pourovskii, T. Miyake, S. I. Simak, A. V. Ruban, L. Dubrovinsky, and I. A. Abrikosov, Physical Review B **87**, 115130 (2013).
21  A. V. Ruban, A. B. Belonoshko, and N. V. Skorodumova, Physical Review B **87**, 014405 (2013).
22  D. F. J. Kimball, et al., Annalen der Physik, n/a (2013).
23  J. M. Brown, S. J. Smullin, T. W. Kornack, and M. V. Romalis, Physical Review Letters **105**, 151604 (2010).